# A Robust UWOC-assisted Multi-hop Topology for Underwater Sensor Network Nodes

Maaz Salman[1], Javad Bolboli[1], and Wan-Young Chung[1*]

[1] Department of AI Convergence, Pukyong National University, Busan 48513, Republic of Korea.

zeesal@pukyong.ac.kr[1], wychung@pknu.ac.kr[4*]

Corresponding author*: Wan-Young Chung

**Abstract** Underwater environment is substantially less explored territory as compared to earth's surface due to lack of robust underwater communication infrastructure. For Internet of Underwater things (IoUT) connectivity, underwater wireless optical communication can play a vital role, compared to conventional radio frequency communication, due to longer range, high data rate, low latency, and unregulated bandwidth. This study proposes underwater wireless optical communication driven local area network (UWOC-LAN), comprised of multiple network nodes with optical transceivers. Moreover, the temperature sensor data is encapsulated with individual authentication identity to enhance the security of the framework at the user end. The proposed system is evaluated in a specially designed water tank of 4 meters. The proposed system's evaluation analysis shows that the system can transmit underwater temperature data reliably in real time. The proposed secure UWOC-LAN is tested within a communication range of 16 meters by incorporating multi-hop connectivity to monitor the underwater environment.

**Keywords**: *Underwater wireless optical communication, local area network, on-off keying*

## 1. INTRODUCTION

In recent years, large number of devices equipped with varieties of sensors have been integrated with the internet combine with the advent of machine learning, realizing the true potential of Internet of Things. This advancement gave boost to range of applications in real world. Furthermore, the concept is reimagined in underwater environments such as ocean, off-shore aquaculture, and ocean-floor monitoring with the coined name, "Internet of Underwater Things (IoUT)" [1]. One of the main applications of IoUT is underwater environment monitoring which collect information such as temperature, humidity, salinity, PH etc. To fully reciprocate the function of IoT in underwater environment, the concept of underwater wireless sensor network (UWSN) has been proposed [2].

Sensor node plays a primary role in UWSN. It comprises of communication modem, battery, and sensors. The sensors collect the environment data and forward to the communication module where the sensed data is transmitted to the gateway (i.e., buoys, ships, or autonomous vehicles). The gateway collected and relayed the information to the nearby monitoring center (via radio communication). Past underwater communication technologies i.e., radio frequency have many challenges such as delay, reliability, mobility, energy consumption, and bandwidth. Therefore, it is challenging to interconnect large amount of UWSN using acoustic and RF technology. To improve the connectivity and number of connected sensor nodes, underwater optical wireless communication (UWOC) can be utilized [3]. UWOC offers wide unregulated frequency spectrum (400-800 THz), low latency, and high data rate [4].

Aquaculture industry is one of the fast-growing food producers in the world. The improvement in the water quality monitoring, management, and the production process of aquaculture can significantly meet the needs of this rapidly growing sector [5]. IoUT can be utilized in aquaculture to automize the process and collect variety of information [6]. Optical communication is considered secure as compared to RF communication by using complicated cryptographic algorithms.

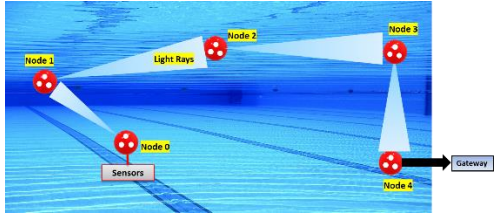

**Fig. 1.** Graphical illustration of multi-hop UWSN.

In this paper, we proposed an end user authentication system to boost the security of the information without using the complicated algorithms. This security-aware framework is difficult to compromise during transmission. UWOC is utilized to interconnect multi-hop nodes to cover an area of 16 meters. The system is equipped with temperature sensor to demonstrate the realizability of aquaculture monitoring. The graphical presentation of the proposed system is illustrated in figure 1. This prototype shows solid capability for future updates. The main contributions of this study are following:

1. We propose a temperature monitoring system utilizing UWOC for underwater environments. The UWSN is consisted of five nodes and each node can sense the temperature of the surrounding environment and transmit it to the monitoring node in real time.
2. The prototype's feasibility and performance are demonstrated via real time experimental evaluation. The achieved results show the feasibility of the proposed system.
3. A security-aware authentication system is proposed, designed, and evaluated to increase the end-to-end encryption of the sensed data.

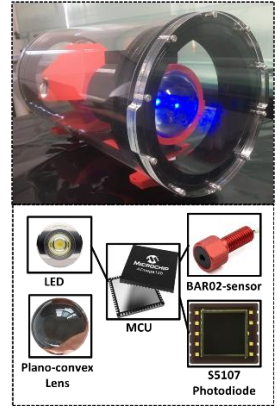

**Fig. 2.** System model of UWOC powered sensor node.

## 2. Underwater Wireless Sensor Network
### 1. Underwater Wireless Sensor Node

The sensor node is comprised of temperature sensor, optical transmitter and receiver module, and collimated lens to aligned and focus the optical beam in a certain direction. The block diagram and the pictorial presentation of the sensor node is shown in figure 2. The temperature sensor (bar02) collects the information and sends to the micro-controller (MCU) Atmega128A where the data is converted to digital signals which drive the optical signal. An LED (Hyper Flux 5pie) of 470 nm wavelength and Thorlabs's plano-convex lens (85 mm focal length) are utilized as an optical source and collimated lens, respectively. The sensor node's algorithm allocates the transmission and reception time during the operation. The photodiode (Hamamatsu S5107) is used to detect the optical signal and realize into meaningful form. A Trans-Impedance Amplifier (Tia) based OPTO-Receiver is used to convert the optical signal into smooth electrical-digital signal [3].

### 2. Security-aware Algorithm

The algorithm receives the transmitted data from the subsequent node, authenticate it using the authentication key of corresponding node and send to the MCU for transmission to the next node. Furthermore, the MCU collects the sensor data, add its authentication key to the data frame along with the received packets and the subsequent node's authentication key. At the final stage, the algorithm calls the transmission function to send the data frame to the following node. The data frame format of each node with their respective authentication key is shown in figure 3.

**Algorithm: Security-aware UWSN**
**Start**
    **Call** Receive function
        | Identify and receive subsequent
        | Sensor nodes' data
    **Call** transmission function
**end**

| Packet Description | Header | Sync Byte | Authentication key | | | Payload | End Byte |
|---|---|---|---|---|---|---|---|
| 1st Node | 255 | 80 | 180 | | | Payload | 00 |
| 2nd Node | 255 | 80 | 180 | 170 | | Payload | 00 |
| 3rd Node | 255 | 80 | 180 | 170 | 154 | Payload | 00 |

**Fig. 3.** Data frame used in the transmission.

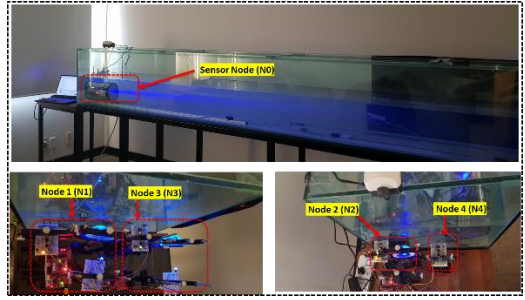

**Fig. 4.** Experimental setup.

**Table 1.** Experimental Parameters

| Parameters | Values |
|---|---|
| Photodiode | Hamamatsu S5107 |
| LED | Hyper flux 5pie blue |
| Lens | Plano-convex |
| Distance | 4-16 meters |
| LED driver chip | LM393 |
| Ambient light | 50-150 lux |

## 3. Experimental Setup

To emulate a relatively larger area for the experiment, a water aquarium of 4 meters is used to conduct the experiment for multi-hop communication. The transmission is performed four times from one end of the aquarium to the other end to cover the link range of 16 meters. Each sensor nodes i.e., node 0, node 1, node 2, node 3, node 4 is connected to the corresponding MCU for data transmission and reception. Each intermediate sensor node is equipped with a pair of optical transmitter and receiver which relay real-time sensor data. Initially, node 0 collects the sensor data and transmits to the following node. The lux intensity of the optical source (LED) is measured after every four meters using lux meter. The system has been evaluated by calculating packet success rate after each transmission (node 0 to node 1, node1 to node 2, node 2 to node 3, node 3 to node 4) with varying turbidities. The turbidity of the water is varied by adding zinc oxide powder in the water. Here, the system is considered as LOS communication and linear topology to demonstrate device-to-device (D2D) communication for the IoT perspective. The graphical illustration of the experimental setup is shown in figure 4. The experimental parameters used in the evaluation of the system is given in table 1.

## 4. Results and Discussions

For performance evaluation, we calculated the packet success rate at the end of each transmission with varying the link distance ( increment of 4 meters) and the turbidity of the water. Moreover, the lux intensity of the optical beam is also measured at each node. Figure 5 demonstrates the packet success rate performance of each transmission. The figure shows comparable lux intensity at each transmission. The system has achieved a packet success rate of about 95 % in 0.01 NTU turbidity level water at the link range of 16 meters whereas 89 % PSR has been achieved in the turbid water of about 70 NTU at the same link range.

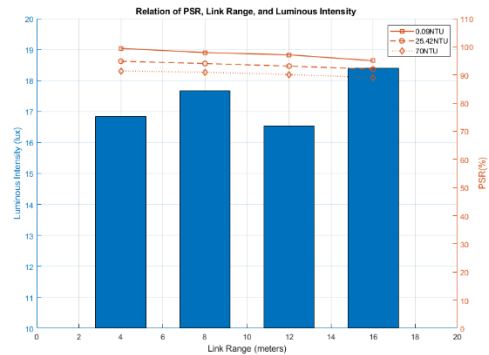

**Fig. 5.** Packet success rate of UWOC system.

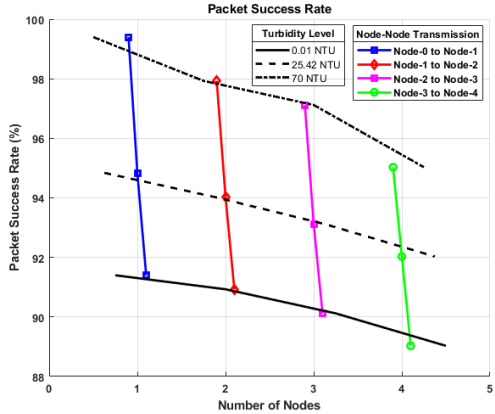

**Fig. 6.** Packet success rate of UWOC system.

Furthermore, the PSR performance of the system is calculated with the addition of hops and change in turbidity levels. The figure 6 shows that after the first transmission (node 0 to node 1), the system has achieved a PSR performance of about 91% in the presence of turbid water of 70 NTU whereas the system has achieved a PSR performance of about 89 %, in the presence of 70 NTU turbid water while transmitting the data from node 3 to node 4. These PSR performances shows that the system is reliable and feasible for underwater environment monitoring for aquaculture and other applications. The monitored temperature data within in span of about few hours has been plotted and illustrated in figure 7.

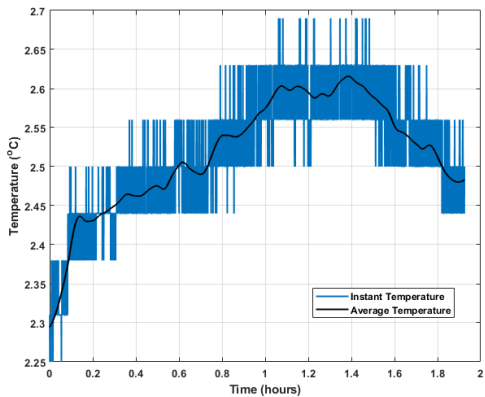

**Fig. 7.** Temperature data.

## 4. CONCLUSIONS

This study proposes a secure multi-hop underwater sensor network using optical wireless communication to interconnect the sensor nodes for underwater environment monitoring. The data is encrypted using an authentication system for each sensor node. The system shows good packet success rate performance in the presence of turbid water of different level.

## ACKNOWLEDGMENTS

This research was funded by National Research Foundation (NRF) of Korea Grant No. 2020R1A4A1019463.